\documentclass[runningheads]{llncs}
\usepackage{graphicx}
\usepackage{booktabs}
\usepackage{threeparttable}
\usepackage{ulem}
\usepackage{color,graphicx}
\usepackage{amsmath}
\usepackage{bbding}

\newcommand{\hollowstar}{\text{\FiveStarOpen}}
\renewcommand{\thanks}[2]{\renewcommand{\thefootnote}{#1}\footnotemark[1]\footnotetext[1]{#2}}

\begin{document}
\title{Anatomy Prior Based U-net for Pathology Segmentation with Attention\thanks{\scriptsize\hollowstar}{This work was funded by the National Natural Science Foundation of China (Grant No. 61971142), and Shanghai Municipal Science and Technology Major Project (Grant No. 2017SHZDZX01).}}
\titlerunning{Anatomy Prior based U-net for Pathology Segmentation}

\author{Anonymous Authors}
\author{Yuncheng Zhou$^\#$, Ke Zhang\thanks{\scriptsize\#}{The authors contributed equally.} \and Xinzhe Luo \and Sihan Wang \and Xiahai Zhuang\thanks*{Xiahai Zhuang is corresponding author. }}
%\authorrunning{Anonymous Authors}
%\institute{Unknown Institutes}
\authorrunning{Y Zhou, et al}
\institute{School of Data Science, Fudan University, Shanghai\\
\email{\{19210980093, 20210980130, zxh\}@fudan.edu.cn}}
\maketitle              % typeset the header of the contribution
\begin{abstract}
Pathological area segmentation in cardiac magnetic resonance (MR) images plays a vital role in the clinical diagnosis of cardiovascular diseases. Because of the irregular shape and small area, pathological segmentation has always been a challenging task. We propose an anatomy prior based framework, which combines the U-net segmentation network with the attention technique. Leveraging the fact that the pathology is inclusive, we propose a neighborhood penalty strategy to gauge the inclusion relationship between the myocardium and the myocardial infarction and no-reflow areas. This neighborhood penalty strategy can be applied to any two labels with inclusive relationships (such as the whole infarction and myocardium, etc.) to form a neighboring loss. The proposed framework is evaluated on the EMIDEC dataset. Results show that our framework is effective in pathological area segmentation.
 
\keywords{Pathology Segmentation  \and Attention Map \and Cardiac MRI}
\end{abstract}
\section{Introduction}

Cardiovascular diseases are the leading cause of death around the world~\cite{huertas2019relevance}. Automated segmentation of cardiac magnetic resonance (CMR) image is an important step towards analyzing cardiac pathology on a large scale, and ultimately the development of diagnosis and treatment methods. Many pioneering works on cardiac MRI segmentation have been done, using traditional machine learning methods or deep learning methods. In \cite{qian2015segmentation,sun2010automatic,zhuang2010registration}, prior knowledge of statistical models is employed to achieve cardiac MRI segmentation. In the last few years, with the rapid improvement of convolutional neural networks (CNNs) in the area of computer vision, they have been proven to be the new state-of-the-arts~\cite{ref_article1,duan2019automatic,ref_article3,ref_article2}. 

Late gadolinium enhancement (LGE) MRI is a sophisticated imaging technique for myocardial infarction (MI) enhancement. Accurate segmentation of the infarction area in medical images has significant prognostic value for infarction diagnosis. However manual segmentation can be time-consuming and suffer from inter-observer heterogeneity. Hence, automated cardiac myocardial infarction segmentation is still desirable. However, the MI segmentation is a challenging task, for example, the effectiveness of the network could be limited due to low soft-tissue contrast, limited training data, irregularity of pathological targets, heterogeneous intensity distributions. Some deep learning-based segmentation methods have been used in the literature to perform MI segmentation~\cite{xu2020segmentation,xu2018mutgan,zhang2018multi}. One prominent method is the deep spatiotemporal adversarial network (DSTGAN). The deep DSTGAN leverages a conditional generative model, which conditions the distributions of the objective CMR image to directly optimize the generalized error of the mapping between the input and the output~\cite{xu2020segmentation}. 

Our uniqueness is that we use U-net with attention and couple the U-net with neighborhood penalty strategy. Different from the conditional generative model-based segmentation framework, our method includes customized attention block \cite{woo2018cbam}. The main contributions of this work are summarized as follows: 

\begin{itemize}
    \item[(1)] We propose a scheme to combine our segmentation network with the attention blocks.
    \item[(2)] We introduce a weight generator to automatically balance the gradient with different labels, the weight generator acts like an assistant and can be plugged to any existing framework (such as U-net, Dense-net, etc).
    \item[(3)] We propose a neighborhood penalty strategy to gauge the myocardium internal pathology inclusion relationship.
\end{itemize}

% \begin{figure}[thb]
% \includegraphics[width=\textwidth]{figure/figure1.pdf}
% \caption{The pipeline of the proposed CMR orientation recognition and standardization method. The image is first truncated at several gray value thresholds. Then the processed image is used to generate the image-orientation pair (see section2.1). Then, the multi-tasking network generates the orientation predicted and segmentation mask. When embed the orientation recognition network to the orientation adjust tool, the multi-tasking network is replaced with a simplified CNN. } \label{fig1}
% \end{figure}

\section{Method}
In this section, we introduce our proposed anatomy prior based U-net for pathology segmentation with attention. The proposed method is based on a deep-learning framework and uses a U-net variant with attention block as the segmentation network. Firstly, we introduce the pathology mix-up augmentation method, which is applied to different myocardium of image pair. Secondly, we present the architecture of the segmentation network with attention block.  Thirdly, we describe the proposed neighborhood penalty strategy and the derived neighboring loss.

\subsubsection{Pathology Mixup}
\begin{figure}[h]
    \centering
    \includegraphics[width=\textwidth]{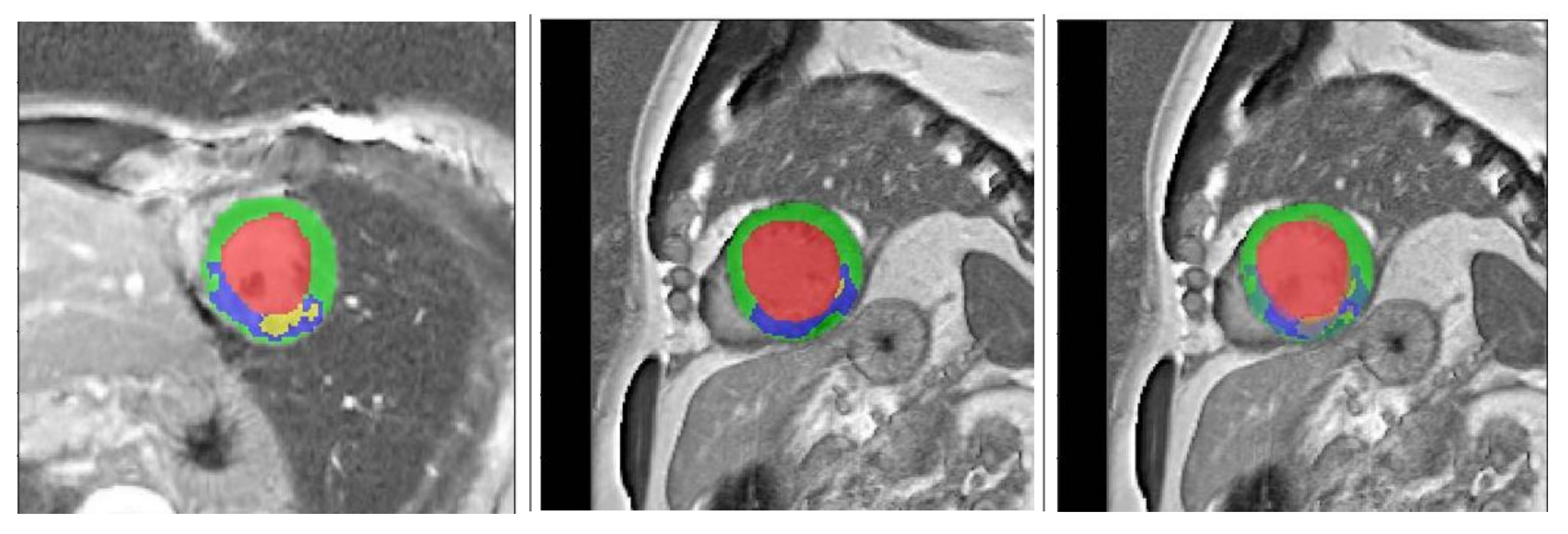}
    \caption{The pathology mix-up process. The middle image is the fixed image, to which left moving image is registered. The right image indicates the mix-up augmentation result.}
    \label{fig:mixup}
\end{figure}

We introduce a data augmentation method particularly for segmentation of cardiac pathology. 
The method is based on the mix-up strategy \cite{zhang2017mixup} to interpolate between two images and their segmentation maps.
Mix-up strategy assume that the corresponding probabilities of a single label are local linear functions of the intensity for a pixel in the images. Hence, in order to let the model be able to simulate such functions, a linearly interpolated probability map for each label was generated for a pair of images. 
Specifically, we randomly select two slices of similar z-positions as long as their segmentation maps.
By treating one as the fixed image $F$ and the other as the moving image $M$, we register the cardiac region in the moving image to the fixed image using affine transformation $T$ including only translation and scaling.
Denote the foreground center (LV + Myocardium) of the fixed and moving image as $c^F=(c^F_x, c^F_y)$ and $c^M=(c^M_x,c^M_y)$, respectively.
The translation offset becomes $\Delta c=(c^F_x-c^M_x, c^F_y-c^M_y)$.
Besides, denote the average distance from foreground pixels to the foreground center as $d^F$ and $d^M$.
The affine matrix we use for affine transformation is
\begin{equation}
    \begin{pmatrix}
      s &\quad 0 &\quad c^F_x - s\cdot c^M_x \\
      0 &\quad s &\quad c^F_y - s\cdot c^M_y \\
      0 &\quad 0 &\quad 1
    \end{pmatrix}
\end{equation}
where $s=d^F/d^M$ is the scaling factor. 
The warped moving image $M\circ T$ and the fixed image $F$ as well as their corresponding transformed segmentation maps are linearly interpolated to create an augmented image-segmentation pair within the foreground area, i.e.
\begin{equation}
    (1-\lambda) M\circ T + \lambda F,
\end{equation}
where $\lambda\sim Uniform(0, 1)$. The background area in the fixed image was used in the augmented image to preserve the outside information of the image. The process is illustrated in Fig. \ref{fig:mixup}.\\[-2em]

\subsubsection{Network Architecture} 
\begin{figure}[h]
\includegraphics[width=\textwidth]{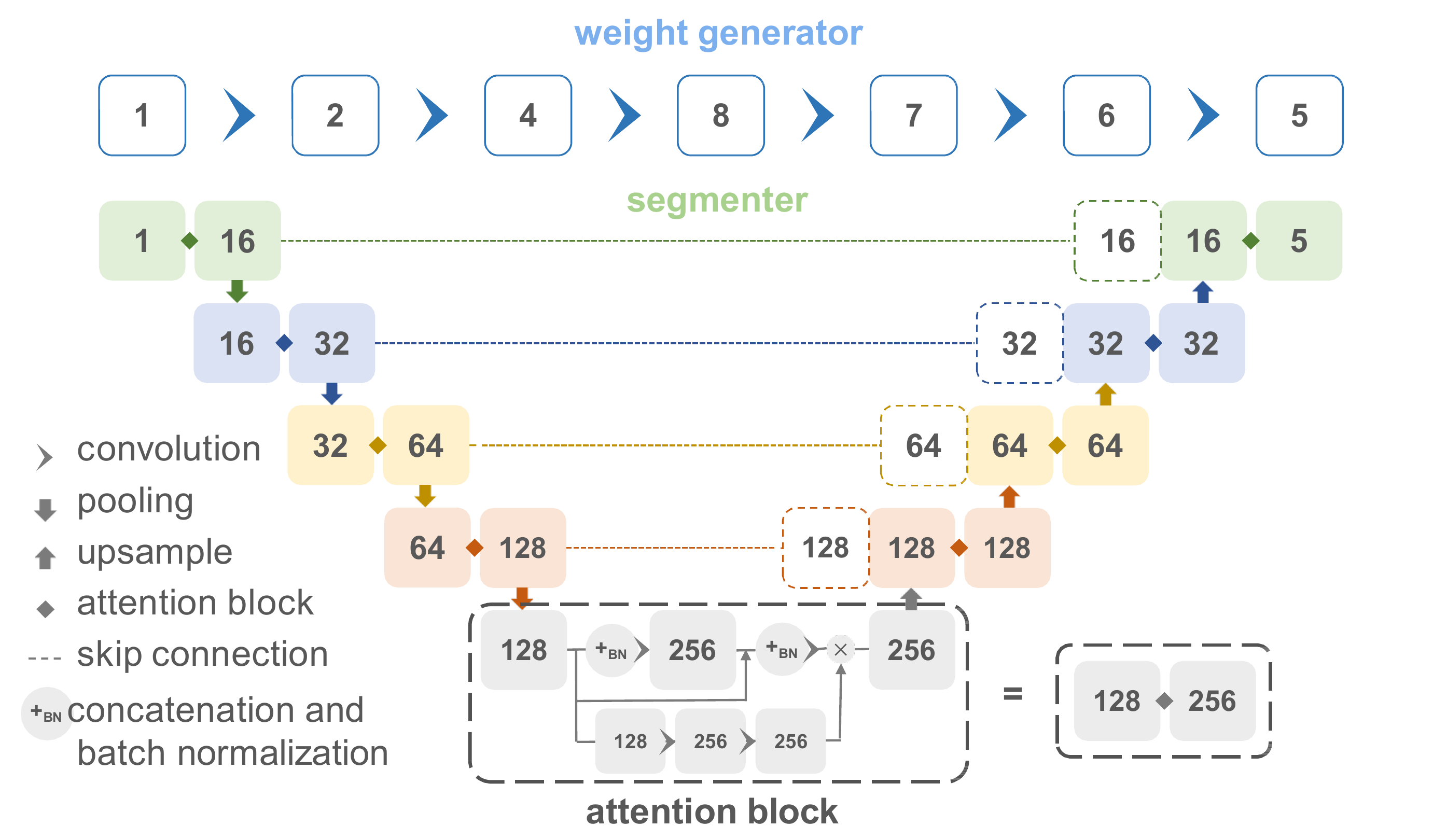}
\caption{The framework of anatomy prior based U-net with attention. Each diamond in the graph represents an attention block while the cross in the attention block means point-wise multiplication.}
\label{fig2}
\end{figure}

We adopt the classical 2D U-net as the backbone. Motivated by the weighted cross-entropy loss which widely used in segmentation tasks, automated weighted cross-entropy loss was used, i.e.
\begin{equation}
L_{AWCE}= -\frac1N\sum_{x}\sum_{i=1}^{s}w_i Y_i(x) \log \hat{Y_i}(x),
\label{eq:wce}
\end{equation}
where $N$ is the number of all pixels, $x$ represents a plixed in the image space,  $i$ stands for the segments while $\hat{Y_i}(x)$ represents the predicted probability for label $i$ and $Y_i(x)$ takes $1$ if the gold standard label of pixel $x$ is $i$ while it is set to $0$ otherwise. To avoid redundant parameter adjustment, we formulate the hyperparameters $w_i$ as an output of the deep neural network. Denote the input augmented image pair as $X_A, Y_A$, the proposed segmentation framework as $S$. $S$ takes as input an image from $X_A$, which is first passed through separate sub-networks, weight generator, $S_{subnet,w}$ and segmentation generator, $S_{subnet,s}$. $S_{subnet,w}$, consists of 6 convolution layers, while $S_{subnet,s}$ consists of U-net encoder and decoder, interleaved with an attention block. The outputs of the two sub-networks are of the same channels. The output of $S_{subnet,w}$ is a weight vector $w$ of dimension 5, which is used to calculate the weighted cross-entropy between the ground truth $Y_A$ and the output segmentation result of $S_{subnet,s}$.  Fig. \ref{fig2} presents the pipeline of the proposed anatomy prior-based segmentation network with attention. 
\\[-3em]

\subsubsection{Neighborhood Penalty Strategy}
Suppose that "the whole" denotes the entire tissue including normal and pathological areas. Leveraging the fact that the no-reflow is contained in the whole infarction, the whole infarction area is contained in the whole myocardium, We propose a neighboring penalty as a weak constraint strategy. Denote $L_{NP}$ as the neighborhood penalty loss, $a$, $b$ as the supporting function for the two labels, which maps the inner part of the region to 1 while the outer part 0. If $a$ and $b$ are not close together, that is $a * b < 1-\varepsilon$ (where $\varepsilon$ is a small given number), the neighboring penalty loss is formulated as:
\begin{equation}
L_{NP}(a,b) = mean(\mathbf 1_{a>0} \times \mathbf 1_{b>0} \times (1-a-b)).
\end{equation}
The penalty encourages two close regions to stick together and thus it prevents cracks. 
In the pathological segmentation of the cardiac MRI, the whole infarction represents the total area of labels for infarction and no-reflow, and the whole myocardium represents the total area of labels for myocardium and the whole infarction. Denote the derived neighboring loss as $L_N$, the no-reflow, infarction, myocardium labels as $Y_{nf}, Y_{in}, Y_{myo}$ respectively, the integral neighboring loss is obtained,
\begin{equation}
L_N = L_{NP}(\hat{Y_{nf}}, \hat{Y_{nf}} + \hat{Y_{in}}) + L_{NP}(\hat{Y_{nf}} + \hat{Y_{inf}}, \hat{Y_{nf}}+\hat{Y_{inf}}+\hat{Y_{myo}}).
\end{equation}

\section{Experiment}

\subsubsection{Materials and implementation}
This experiment was performed on the \newline EMIDEC challenge dataset \cite{emidec}, which comprises 150 cardiac MRI images from different patients including 50 cases with normal MRI and 100 cases with myocardial infarction, of which 100 subjects were randomly selected as the training dataset with 67 cases of pathological cases and 33 normal cases. Test data are composed of 50 other patients with 33 pathological cases and 17 normal cases.

In our experiments, the proposed models were trained on 90\% of training slices while the other 10\% was chosen to be the validation set. As a result, 639 slices are used as training data while the other 69 slices are used to evaluate the model.

For the proposed loss terms, neighborhood penalty was provided with a coefficient of $10^{-2}$ while the entropy loss's coefficient was always set to $1$ with or without AWCE applied. 

All experiments were conducted using the PyTorch framework with Python 3.7.4. For each iteration, a mini-batch of size 16 was fed to the network and the parameters were trained for 400 epochs. To accelerate the training process and avoid convergence to local minima, an Adam optimizer with a decreasing learning rate from $10^{-2}$ to $10^{-6}$ was used. The training process was performed on a NVIDIA GTX1080Ti GPU. 

\subsubsection{Ablation study}
Under a running environment with previously mentioned hard and software, a dense U-net was trained and evaluated. This model was designed after the proposed segmentation generator shown in Fig. \ref{fig2}. In this network, all auxiliary attention branches were removed from the proposed architecture. By using cross-entropy as the energy function, the optimal model was regarded as the baseline model. 

A step-by-step increment of the algorithm was performed as an ablation study to prove the effectiveness of the implanted modules and techniques. 
First, we regard the attention mechanism as a good way to improve the segmentation of small and sensitive regions such as the infarcted areas. Specifically speaking, two convolutions were applied to the input feature maps of the U-net blocks and the output was normalized between $0$ and $1$ by the sigmoid function. This normalized output was then regarded as used to activate the output feature maps. Secondly, the automated weighted cross-entropy (equation \ref{eq:wce}) was used to automatically balance the gradient caused by different labels. The trained weight generator acts as an assistant to the main network but it does not affect the evaluation process. Thirdly, the neighborhood penalty was also added to the method to regularize the segmentation by shape prior. The coefficient for this penalty was set to $0.01$ when the magnitude of the penalty is similar to that of the entropy loss.

\subsubsection{Results and discussion}

\begin{figure}[thb]
\includegraphics[width=\textwidth]{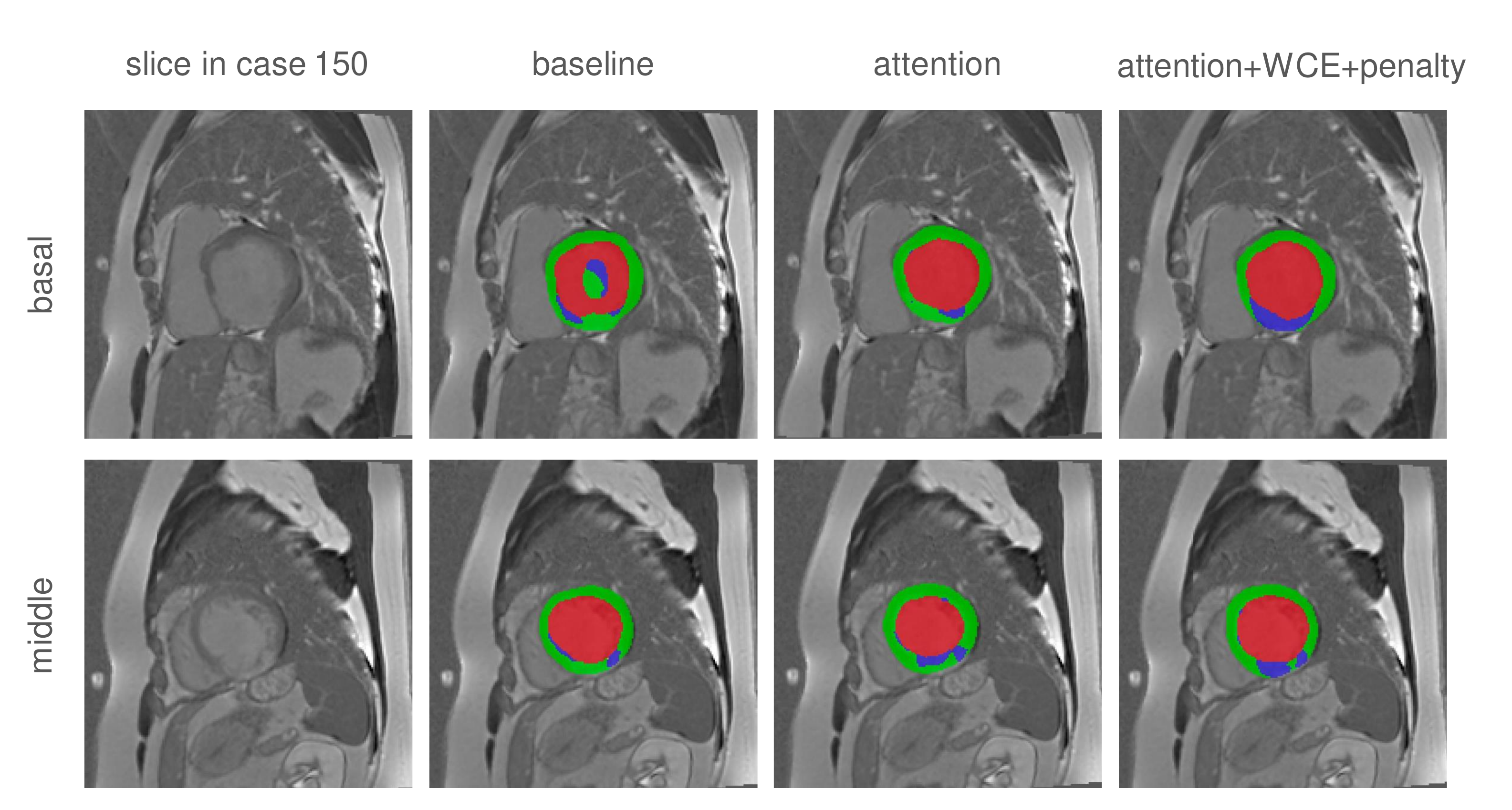}
\caption{Qualitative comparison of the ablation study experiments} 
\label{fig3}
\end{figure}

Table \ref{table1} presents quantitative results of our experiments. The reported numbers are the mean Dice score coefficient (DSC) and the standard deviation on the validation set. The improvements can also be visually observed in figure \ref{fig3}. In the rest of this section, we discuss the results of the ablation study experiments. It can be observed that the proposed attention block method provides substantial improvements over dense U-net baseline. The proposed generator based weighted cross entropy method improves performance as compared to the U-net with attention method. These results show the benefit of encouraging the automatic label weighting to solve the label imbalance problem.

The no-reflow and infarcted areas are commonly difficult to be segmented due to its blurry boundary and its small area. However, the design of the weighted entropy solved the problem of small area while the attention blocks helped to refine the detailed segmentation boundary by decreasing the propagated gradients of unwanted areas. These techniques, as expected, improve the results. 

As an additional experiment, we investigated the effect of including the neighboring loss from neighborhood penalty strategy. While the neighborhood penalty loss resulted in worse performance than the generator based weighted cross-entropy in general, it provides improvements for the most difficult no-reflow label. This shows that although the neighboring loss does not ensure the learning of accurate shape prior, it is still advantageous to enforce soft inclusive constraints on anatomy labels.

\begin{table*}[h]
\centering
\caption{DSC of segmentation on the validation slices. In this table, LV stands for left ventrical, Myo stands for myocardium while Inf and NoR stand for the infarcted area and no-reflow area respectively. }
\label{tab:result}
\begin{tabular}{l|cccc}
	\bottomrule
	&LV&Myo&Inf&NoR\\
	\hline
	baseline&
	$0.955\pm0.009$&$0.871\pm0.045$&$0.622\pm0.080$&$0.246\pm0.102$\\
	\hline
	attention&
	$0.962\pm0.009$&$0.900\pm0.033$&$0.718\pm0.067$&$0.354\pm0.130$\\
	attention+AWCE+penalty&
	$0.970\pm0.007$&$0.916\pm0.029$&$0.747\pm0.082$&$\mathbf{0.538\pm0.143}$\\
	attention+AWCE&
	$\mathbf{0.971\pm0.014}$&$\mathbf{0.926\pm0.029}$&$\mathbf{0.769\pm0.082}$&$0.535\pm0.153$\\
	\toprule
\end{tabular}\label{table1}
\end{table*}

\section{Conclusion}
We have proposed an anatomy prior based framework that consists of a weight generator and segmentation generator with attention block. Also, we have presented a neighborhood penalty strategy, which measures the inclusive relationship and acts as a weak constraint. The experiment demonstrates that the proposed framework can segment cardiac MRI images with pathology effectively. Our future research aims to optimize the augmentation methods and investigate our proposed framework in diverse training scenarios.
\bibliographystyle{splncs04}
\bibliography{paper27.bib}

\begin{thebibliography}{10}
\providecommand{\url}[1]{\texttt{#1}}
\providecommand{\urlprefix}{URL }
\providecommand{\doi}[1]{https://doi.org/#1}

\bibitem{ref_article1}
Bernard, O., Lalande, A., Zotti, C., Cervenansky, F., Yang, X., Heng, P.A.,
  Cetin, I., Lekadir, K., Camara, O., Ballester, M.A.G., et~al.: Deep learning
  techniques for automatic mri cardiac multi-structures segmentation and
  diagnosis: is the problem solved? IEEE transactions on medical imaging
  \textbf{37}(11),  2514--2525 (2018)

\bibitem{duan2019automatic}
Duan, J., Bello, G., Schlemper, J., Bai, W., Dawes, T.J., Biffi, C., de~Marvao,
  A., Doumoud, G., O’Regan, D.P., Rueckert, D.: Automatic 3d bi-ventricular
  segmentation of cardiac images by a shape-refined multi-task deep learning
  approach. IEEE transactions on medical imaging  \textbf{38}(9),  2151--2164
  (2019)

\bibitem{huertas2019relevance}
Huertas-Vazquez, A., Leon-Mimila, P., Wang, J.: Relevance of multi-omics
  studies in cardiovascular diseases. Frontiers in cardiovascular medicine
  \textbf{6}, ~91 (2019)

\bibitem{emidec}
Lalande, A., Chen, Z., Decourselle, T., Qayyum, A., Pommier, T., Lorgis, L.,
  la~Ezequiel, R., Cochet, A., Cottin, Y., Ginhac, D., Salomon, M., Couturier,
  R., Meriaudeau, F.: Emidec: A database usable for the automatic evaluation of
  myocardial infarction from delayed-enhancement cardiac mri. Data 2020
  \textbf{5}(4), ~89 (2020)

\bibitem{qian2015segmentation}
Qian, X., Lin, Y., Zhao, Y., Wang, J., Liu, J., Zhuang, X.: Segmentation of
  myocardium from cardiac mr images using a novel dynamic programming based
  segmentation method. Medical physics  \textbf{42}(3),  1424--1435 (2015)

\bibitem{sun2010automatic}
Sun, H., Frangi, A.F., Wang, H., Sukno, F.M., Tobon-Gomez, C., Yushkevich,
  P.A.: Automatic cardiac mri segmentation using a biventricular deformable
  medial model. In: International Conference on Medical Image Computing and
  Computer-Assisted Intervention. pp. 468--475. Springer (2010)

\bibitem{woo2018cbam}
Woo, S., Park, J., Lee, J.Y., So~Kweon, I.: Cbam: Convolutional block attention
  module. In: Proceedings of the European conference on computer vision (ECCV).
  pp. 3--19 (2018)

\bibitem{xu2020segmentation}
Xu, C., Howey, J., Ohorodnyk, P., Roth, M., Zhang, H., Li, S.: Segmentation and
  quantification of infarction without contrast agents via spatiotemporal
  generative adversarial learning. Medical Image Analysis  \textbf{59},  101568
  (2020)

\bibitem{xu2018mutgan}
Xu, C., Xu, L., Brahm, G., Zhang, H., Li, S.: Mutgan: Simultaneous segmentation
  and quantification of myocardial infarction without contrast agents via joint
  adversarial learning. In: International Conference on Medical Image Computing
  and Computer-Assisted Intervention. pp. 525--534. Springer (2018)

\bibitem{zhang2018multi}
Zhang, D., Icke, I., Dogdas, B., Parimal, S., Sampath, S., Forbes, J., Bagchi,
  A., Chin, C.L., Chen, A.: A multi-level convolutional lstm model for the
  segmentation of left ventricle myocardium in infarcted porcine cine mr
  images. In: 2018 IEEE 15th International Symposium on Biomedical Imaging
  (ISBI 2018). pp. 470--473. IEEE (2018)

\bibitem{zhang2017mixup}
Zhang, H., Cisse, M., Dauphin, Y.N., Lopez-Paz, D.: mixup: Beyond empirical
  risk minimization. arXiv preprint arXiv:1710.09412  (2017)

\bibitem{ref_article3}
Zhuang, X.: Multivariate mixture model for cardiac segmentation from
  multi-sequence mri. In: International Conference on Medical Image Computing
  and Computer-Assisted Intervention. pp. 581--588. Springer (2016)

\bibitem{ref_article2}
Zhuang, X.: Multivariate mixture model for myocardial segmentation combining
  multi-source images. IEEE transactions on pattern analysis and machine
  intelligence  \textbf{41}(12),  2933--2946 (2019)

\bibitem{zhuang2010registration}
Zhuang, X., Rhode, K.S., Razavi, R.S., Hawkes, D.J., Ourselin, S.: A
  registration-based propagation framework for automatic whole heart
  segmentation of cardiac mri. IEEE transactions on medical imaging
  \textbf{29}(9),  1612--1625 (2010)

\end{thebibliography}

\end{document}